# Use of L-system mathematics for making new subfamily members of olfactory receptor full length genes, OR1D2, OR1D4 and OR1D5.

Sk. Sarif Hassan[a,c], Pabitra Pal Choudhury[a], Amita Pal[b], R. L. Brahmachary[c] and Arunava Goswami[c,1]

[a]Applied Statistics Unit, Indian Statistical Institute, 203 B. T. Road, Calcutta, 700108 India. pabitrapalchoudhury@gmail.com [P. P. C.]; sarimif@gmail.com [S. S. H.];

[b] Bayesian Interdisciplinary Research Unit (BIRU), Indian Statistical Institute, 203 B. T. Road, Calcutta, 700108 India. pamita@isical.ac.in [A. P.] and

[c]Biological Sciences Division, Indian Statistical Institute, 203 B. T. Road, Calcutta, 700108 India. agoswami@isical.ac.in [A.G.].

**Keywords**
- Human olfactory receptor
- L-system
- ClustalW
- Star Model
- Olfaction

**Footnotes**
- [1]To whom correspondence should be addressed. E-mail: agoswami@isical.ac.in / srabanisopanarunava@gmail.com
- *Author contributions*: A. G. conceptualized the experiments and performed entire research with S. S. H.; A. G. and S. S. H. acknowledges A. P., P. P. C. and R. L. B. and Ms. Rebecca French (Grad student of Prof. Michael Hochella) of Virginia Tech, USA for their enormous help while doing these series of experiments.
- Conflict of interest statement: The authors declare no conflict of interest.

**Abstract**

Ligands for only two human olfactory receptors are known. One of them, OR1D2, binds to Bourgeonal [Malnic B, Godfrey P-A, Buck L-B (2004) The human olfactory receptor gene family. *Proc. Natl. Acad. Sci U. S. A*. 101: 2584-2589 and Erratum in: *Proc Natl Acad Sci U. S. A*. (2004) 101: 7205]. OR1D2, OR1D4 and OR1D5 are three full length olfactory receptors present in an olfactory locus in human genome. These receptors are more than 80% identical in DNA sequences and have 108 base pair mismatches among them. We have used L-system mathematics and have been able to show a closely related subfamily of OR1D2, OR1D4 and OR1D5.

**Introduction**

Olfactory receptors (ORs) loci in human genome occur in clusters ranging from ~51-105 and they are unevenly spread over 21 chromosomes (1, 2). A conservative estimate suggests that 339 full length OR genes and 297 OR pseudogenes are present in these clusters (1). Theoretically, there are two possible ways of OR-odorant molecular binding, viz., (i) each OR binds to a large number of different odorants and (ii) each OR binds to a small number of odorants. In either case, odorant detection at the OR level follows a combinatorial rule, though the stringency of the rule would differ in the two alternatives. Experimentally, it has been demonstrated that each OR recognizes a large number of odorants and perhaps a large class of various concentrations of the odorants tested (3). OR gene (conceptually translated to protein sequences) family (>40% amino acid identity) can be divided into subfamily (>60% identity) and sub-family members might have more than 90% identity (4). Subfamily members are highly similar in DNA and protein sequences, but they are capable of recognizing different odorant molecules.

Three full length model subfamily OR members from HORDE database (http://genome.weizmann.ac.il/horde/), OR1D2 (Gene length: 936 bp), OR1D4 (Gene length: 936 bp) and OR1D5 (Gene length: 936 bp) were downloaded from the HORDE database. OR1D2, OR1D4 and OR1D5 were aligned using ClustalW and was found to contain 108 base pair mismatches out of 936 base pairs available (data not shown). OR1D2, OR1D4 and OR1D5 are highly related sequences, therefore, a canonical sequence for this subfamily, termed as `star model' of OR sequence was made by using a computer C program, where 108 gaps were introduced in respective positions following a computer algorithm given at the end of the paper in reference (Fig 1a and Fig 1b).

A context free L-system (5), was used to generate a 243 bp long DNA sequence.

**L System:**

**Set of Variables:** A, T, C, and G.

**Axiom:** C (C is the starting symbol)

**Production Rule**:    A → CTG, C→CCA, T→TGC and G→GAC

Following aforesaid production rule, 1$^{st}$ and 2$^{nd}$ iteration, would give CCA (03 bp) and CCACCACTG (09 bp) respectively. Four iterations yield 81 base pair sequences. This is insufficient to answer for 108 mismatches. Five such iterations generate the following 243 bp sequence-

CCACCACTGCCACCACTGCCATGCGACCCACCACTGCCACCACTGCCATGCGACCCACCACT
GTGCGACCCAGACCTGCCACCACCACTGCCACCACTGCCATGCGACCCACCACTGCCACCAC
TGCCATGCGACCCACCACTGTGCGACCCAGACCTGCCACCACCACTGCCACCACTGCCATGC
GACTGCGACCCAGACCTGCCACCACCACTGGACCTGCCACCATGCGACCCACCACTG…(i)

Using a C computer program, nucleotides present in sequence (i) was introduced from 5'- end of the sequence into the star model gaps shown in Fig. 1 sequentially. Briefly, the

Step 1: First, in all the gaps (with 1 bp, 2 bp, 3 bp and 4 bp) in star model, only one nucleotide would be inserted.

Step 2: 1 bp gaps in star model would become 0 gap. Then in the remaining gaps (1 bp, 2 bp and 3 bp) would be filled up and the process would be repeated until all gaps are filled.

The resultant OR sequence is shown in (ii) below.

ATGGATGGAGCCAACCAGAGTGAGTCCTCACAGTTCCTTCTCCTGGGGATGTCAGAGAGTCC
TGAGCAGCAGCAGATCCTGTTTTGGATGTTCCTGTCCATGTACCTGGTCACGGTGCTGGGAA
ATGTGCTCATCATCCTGGCCATCAGCTCTGATTCCCCCCTGCACACCCCCGTGTACTTCTTCC
TGGCCAACCTCTCCTTCACTGACCTCTTCTTTGTCACCAACACAATCCCCAAGATGCTGGTGA
ACCTCCAGTCCCAGAACAAAGCCATCTCCTATGCAGGGTGTCTGACACAGCTCTACTTCCTG
GTCTCCTTGGTGACCCTGGACAACCTCATCCTGGCCGTGATGGCCTATGATCGCTATGTGGCC
AGCTGCTGCCCCCTCCACTACGCCACAGCCATGAGCCCTGCGCTCTGTCTCTTCCTCCTGTCC
TTGTGTTGGGCGCTGTCAGTCCTCTATGGCCTCCTGCCCACCGTCCTCATGACCAGCGTGACC
TTCTGTGGGCCTCGAGACATCCACTACGTCTTCTGTGACATGTACCTGGTGCTGCGGTTGGCA
TGTTCCAACAGCCACATGAATCACACAGCGCTGATTGCCACGGGCTGCTTCATCTTCCTCACT
CCCTTGGGATTCCTGACCAGGTCCTATGTCCCCATTGTCAGACCCATCCTGGGAATACCCTCC
GCCTCTAAGAAATACAAAGCCTTCTCCACCTGTGCCTCCCATTTGGGTGGAGTCTCCCTCTTA
TATGGGACCCTTCCTATGGTTTACCTGGAGCCCCTCCATACCTACTCCCTGAAGGACTCAGTA
GCCACAGTGATGTATGCTGTGGTGACACCCATGATGAACCCGTTCATCTACAGCCTGAGGAA
CAAGGACATGCATGGGGCTCAGGGAAGACTCCTACGCAGACCCTTTGAGAGGCAAACA (ii)

Sequence (ii) was blasted using DNA-DNA and translated protein-protein (Blastx) search engines in HORDE and NCBI database from where initial OR1D2, OR1D4 and OR1D5 sequences were obtained. Results of blast searches show that with the search parameters available in the HORDE website (which could not be changed by remote user), the (ii) sequence showed 92%, 92% and 91% identity with OR1D2, OR1D4 and OR1D5 respectively. Significantly, these insertions do not produce any stop codon in the exon sequence. Therefore, it is clear from the above result that close relative of OR1D2, OR1D4 and OR1D5 subfamily can be generated by this approach. In the next paper, we have shown that L-system can also be used to generated for generating close relative of a single pseudogene present in the same loci that of OR1D2, OR1D4 and OR1D5.

In summary, in this paper, we report that close relatives of OR1D2, OR1D4 and OR1D5 can be generated by using L-system mathematics.

## Appendix

While writing the computer programme, L-System satisfied following rules- A. To fill the <u>single</u> gap of the Star Model: System will check two previous states as well as the two past states of the gap- (a) If the second previous is 'T' and the first previous is 'A' and the first past is 'A' and the second past is either 'A' or 'G', then the chosen L-System must produce 'C' at the gap. e.g., …TA_AA(/G)… (b) If the second previous is 'T' and the first previous is 'A' and the first past is 'G' and the second past is 'A', then the chosen L-System must produce 'C' at the gap. e.g., …TA_GA… (c) If the second previous is 'T' and the first previous is 'A' and the first past is either 'T' or 'C', then the chosen L-System must produce 'C' or 'T' at the gap. e.g., …TA_T(/C)… (d) If the second previous is 'T' and the first previous is 'G' and the first past is 'A' and the second past is either 'A' or 'G', then the chosen L-System must produce 'C' or 'G' at the gap. e.g., …TG_AA(/G)… (e) If the second previous is 'T' and the first previous is 'G' and the first past is 'G' and the second past is 'A', then the chosen L-System must produce 'C' or 'G' at the gap. e.g., …TG_GA… (f) If the second previous is 'T' and the first previous is 'G' and the first past is either 'T' or 'C', then the chosen L-System must produce 'C' or 'G' or 'T' at the gap. e.g., …TG_T(/C)… (g) If the first previous is 'T' and the first past is 'A' and the second past is either 'C' or 'T', then the chosen L-System must produce 'T' or 'C' at the gap. e.g., …T_AC(/T)… (h) If the first previous is 'T' and the first past is 'A' and the second past is either 'A' or 'G', then the chosen L-System must produce 'C' at the gap. e.g., …T_AA(/G)… (i) If the first previous is 'T' and the first past is 'G' and the second past is 'A', then the chosen L-System must produce 'G' or 'C' at the gap. e.g., …T_GA… (j) If the first previous is 'T' and the first past is 'G' and the second past is 'C' or 'T' or 'G', then the chosen L-System must produce 'T' or 'C' or 'G' at the gap. e.g., …T_GC(/T/G)… (h) If the first previous is 'C' and the first past is 'A' and the second past is either 'A' or 'G', then the chosen L-System must produce 'C' or 'G' or 'A' at the gap. e.g., …C_AA(/G)… (i) If the first previous is 'C' and the first past is 'G' and the second past is 'A', then the chosen L-System must produce 'C' or 'G' or 'A' at the gap. e.g., …C_GA… (j) Else the gap can be filled by any state like 'A' or 'C' or 'T' or 'G'. B. To fill the double or more than double gap of the Star Model (one gap fill at a time): System should check only two previous states of the gap- (a) If the second previous is 'T' and the first previous is 'A', then the chosen L-System must produce 'C' or 'T' at the gap. e.g., …TA_ _… (b) If the second previous is 'T' and the first previous is 'G', then the chosen L-System must produce 'C' or 'T' or 'G' at the gap. e.g., …TG_ _… (c) Else the gap can be filled by any state like 'A' or 'C' or 'T' or 'G'. This rule would be applicable until the number of gap becomes one. When the no. of gap becomes one, then the rule (A) is applicable.

## Acknowledgements

This work was supported by Department of Biotechnology (DBT), Govt. of India grants (BT/PR9050/NNT/28/21/2007 & BT/PR8931/NNT/28/07/2007 to A. Goswami) & NAIP-ICAR-World Bank grant (Comp-4/C3004/2008-09; Project leader: A. Goswami) and ISI plan projects for 2001-2011. Authors are grateful to their visiting students Rajneesh Singh, Snigdha Das and Somnath Mukherjee for their technical help in making advanced C programs and other computer applications on Windows support used for this study.

## References

1. Malnic B, Godfrey P-A, Buck L-B (2004) The human olfactory receptor gene family. *Proc. Natl. Acad. Sci U. S. A.* 101: 2584-2589 and Erratum in: *Proc Natl Acad Sci U. S. A.* (2004) 101: 7205.

2. Young J-M, Endicott R-M, Parghi S-S, Walker M, Kidd J-M, Trask B-J (2008) Extensive copy-number variation of the human olfactory receptor gene family. *Am. J. Hum. Genet.* 83: 228–242.

3. Malnic B, Hirono J, Sato T, Buck L-B (1999) Combinatorial receptor codes for odors. *Cell* 96: 713-723.

4. Glusman G, Yanai I, Rubin I, Lancet D (2001) The complete human olfactory subgenome. *Genome Research* 11: 685-702.

5. Prusinkiewicz P, Lindenmayer A (1990) in *The algorithmic beauty of plants*, Springer-Verlag ISBN 978-0387972978.